\documentclass[12pt]{article}

\usepackage{amsfonts}
\usepackage{cite}

\begin{document}

\newcommand{\PT}{{\cal P}{\cal T}}

\title{\large\bf A  Bethe ansatz solvable model for 
superpositions of Cooper pairs and condensed molecular bosons}

\author{K. E. Hibberd$^\spadesuit$, C. Dunning$^\diamondsuit $  and J. Links$^\spadesuit$  \\
$^\spadesuit $Centre for Mathematical Physics, \\
The University of Queensland, 4072, Australia   \\
$^\diamondsuit $Institute of Mathematics, Statistics 
 and Actuarial Science \\
The University of Kent, U.K. \\
  }

\maketitle

\vspace{10pt}

\begin{abstract}
We introduce a general Hamiltonian describing coherent superpositions of Cooper pairs and condensed molecular bosons. 
For particular choices of the coupling parameters, the model is integrable. 
One integrable manifold, as well as the Bethe ansatz solution, was found by 
{\it Dukelsky et al., Phys. Rev. Lett. {\bf 93} (2004) 050403}. 
Here we show that there is a second integrable manifold, 
established using the boundary Quantum Inverse
Scattering Method. In this manner we obtain the exact solution by means of the
algebraic Bethe ansatz.  
In the case where the Cooper pair energies are degenerate 
we examine the relationship between the spectrum of these 
integrable Hamiltonians and 
the quasi-exactly solvable spectrum of 
particular Schr\"odinger operators.   
For the solution we derive here the potential of the Schr\"odinger operator 
is given in terms of hyperbolic functions. For the solution derived by
{\it Dukelsky et al., loc. cit.}  the potential is sextic 
and the wavefunctions obey 
$\PT$-symmetric boundary conditions. This latter case provides a novel example of an integrable Hermitian Hamiltonian acting on a Fock space whose states map into a Hilbert space of $\PT$-symmetric wavefunctions defined on a contour in the complex plane.    

\end{abstract}

\newcommand{\ot}{\otimes}
\newcommand{\tmod}{{\cal T}}
\newcommand{\amod}{{\cal A}}
\newcommand{\bemod}{{\cal B}}
\newcommand{\cmod}{{\cal C}}
\newcommand{\dmod}{{\cal D}}
\newcommand{\hmod}{{\cal H}}
\newcommand{\einsop}{{\bf 1}}
\def\oR{R^*} 
\def\upa{\uparrow}
\def\R{\overline{R}} 
\def\doa{\downarrow}
\def\oL{\overline{\Lambda}}
\def\nn{\nonumber} 
\def\dag{\dagger}
\def\beq{\begin{equation}}
\def\eeq{\end{equation}}
\def\bea{\begin{eqnarray}}
\def\eea{\end{eqnarray}}
\def\ve{\varepsilon}
\def\w{\overline{w}}
\def\u{\overline{u}}
\def\rr{\mathcal{R}}
\def\T{\mathcal{T}}
\def\N{\overline{N}}
\def\Q{\overline{Q}}
\def\i{\overline{i}}
\def\j{\overline{j}}
\def\k{\overline{k}}
\def\de{\delta} 
\newcommand{\reff}[1]{eq.~(\ref{#1})}

\def\d{\delta}
\def\a{\alpha}
\def\b{\beta}
\def\da{\dagger}
\def\e{\epsilon}
\def\g{\gamma}
\def\k{\kappa}
\def\l{\lambda}
\def\o{\omega}
\def\O{\Omega}
\def\D{\Delta}
\def\T{{\cal T}}
\def\TT{{\tilde{\cal T}}}

\def\ba{\begin{array}}
\def\ea{\end{array}}
\def\no{\nonumber}
\def\le{\langle}
\def\re{\rangle}
\def\lt{\left}
\def\rt{\right}
\def\oL{\overline{\Lambda}}
\def\s{\sigma}
\def\t{\theta}

\def\w{\overline{w}}
\def\u{\overline{u}}
\def\rr{\mathcal{R}}
\def\T{\mathcal{T}}
\def\N{\overline{N}}
\def\Q{\overline{Q}}
\def\i{\overline{i}}
\def\j{\overline{j}}
\def\k{\overline{k}}
\newcommand{\cte}[1]{[{\small\bf #1}] }
\newcommand{\fract}[2]{{\textstyle\frac{#1}{#2}}}
\newcommand{\fr}[2]{{\textstyle\frac{#1}{#2}}}
\newcommand{\sech}{{\ \rm sech}}
\newcommand{\ep}{\varepsilon}
\newcommand{\CS}{{\cal S}}
\newcommand{\CL}{{\cal L}}
\newcommand\ZZ{{\mathbb Z}}
\newcommand\RR{{\mathbb R}}
\newcommand\NN{{\mathbb N}}
\newcommand\AAa{{\mathbb A}}
\newcommand\HH{{\mathbb H}}
\newcommand\MM{{\mathbb M}}
\newcommand\Zt{{\mathbb Z}_2}
\newcommand\Zth{{\mathbb Z}_3}
\newcommand\Zf{{\mathbb Z}_5}
\newcommand\Zs{{\mathbb Z}_6}
\newcommand\ZN{{\mathbb Z}_N}
\newcommand\Zn{{\mathbb Z}_n}
\newcommand{\balpha}{\alpha\kern -6.7pt\alpha}
\newcommand{\bbalpha}{\alpha\kern -4.95pt\alpha}
\newcommand\phup{^{\phantom p}}
\newcommand{\CA}{{\cal A}}
\newcommand{\CaC}{{\cal C}}
\newcommand{\CH}{{\cal H}}
\newcommand{\CM}{{\cal M}}
\newcommand{\CN}{{\cal N}}
\newcommand{\CK}{{\cal K}}
\newcommand{\CQ}{{\cal Q}}
\newcommand{\CQJ}{{\cal Q}\phup_J}
\newcommand{\bg}{{\bf g}}
\newcommand{\Hs}{{\CH}^{(2)}}
\newcommand{\Ht}{{\CH}^{(3)}}
\newcommand\half{{\textstyle\frac{1}{2}}}
\newcommand\hf{\fract{1}{2}}
\newcommand\Epsilon{{\cal E}}
\newcommand\Ai{{\rm Ai}}
\newcommand{\wt}{\widetilde}
\newcommand{\FT}[1]{{\cal F}\!\left[#1\right]}
\def\w{\omega}
\newcommand{\gam}{\gamma}
\newcommand{\bl}{ \bar{l}}
\newcommand{\tl}{\tilde{l}}
\newcommand{\eff}{\rm eff}
\newcommand{\One}{{\hbox{{\rm 1{\hbox to 1.5pt{\hss\rm1}}}}}}
\renewcommand{\One}{{\mathbb 1}}
\renewcommand{\One}{{\rm 1\!\!1}}
\newcommand{\sss}[1]{\scriptscriptstyle{\rm #1}}
\newcommand\MR{M\!R}
\newcommand{\widetable}{\renewcommand{\arraystretch}{1.1}}
\newcommand{\ceff}{ c_{\rm eff}}



\def\dwn{\downarrow}
\def\up{\uparrow}
\def\dag{\dagger}
\def\nn{\nonumber}
\def\L{\Lambda}

\vskip.3in


\section{Introduction\label{int}}


One of the most interesting developments in the physics of ultra-cold gases has been 
the production of molecular condensates
from fermionic constituents \cite{grj,rgj}. Due to the property of Pauli blocking, it turns out 
that molecular condensates formed from fermionic atoms are generally more resistant to 
decay compared to the case of molecules formed from bosonic atoms. Furthermore, such 
systems open possibilities to experimentally probe the nature of the crossover between BCS and BEC type physics \cite{pss}. 

In order to theoretically model such systems, it is useful to have
integrable systems  
with exact solutions which allow for 
investigations beyond the limits of perturbative and mean-field treatments. Recently, 
Dukelsky et al. \cite{ddep,duk} have proposed
such a model and derived an exact solution using the methods of Richardson-Gaudin type 
spin models \cite{dps}. In their approach, a bosonic degree of freedom describing the molecular 
condensate is introduced by taking the infinite spin limit of one realisation of particular 
spin operators following a procedure of Gaudin \cite{gaudin}. The remaining spin operators are then 
realised in terms of paired creation and annihilation operators, which may be either bosonic or fermionic. 

The aim of the present work is to derive a similar, although significantly distinct, model which 
also admits an exact solution. 
The approach we adopt is to use the boundary Quantum Inverse Scattering Method (QISM) as developed 
by Sklyanin \cite{skyl}. We construct a doubled monodromy matrix from the spin realisation of the 
Yang--Baxter algebra in the usual way 
but with a bosonic operator valued solution of the reflection equations, which is obtained by dressing 
a boundary $K$-matrix with the bosonic realisation of the Yang--Baxter algebra. For 
simplicity, here we only consider the case where the spin operators are realised in terms of Cooper 
pairs of spin 1/2 particles, although generalisations to higher spin particles are possible. 
Through this construction we yield a family of commuting transfer matrices which give rise to an 
abstract integrable system. Following the method used to establish the integrability of the Russian Doll BCS model 
\cite{dl}, we obtain the conserved charges by expanding the transfer matrix in inverse powers of the  
spectral parameter. Finally, by taking the quasi-classical limit 
(e.g. see \cite{jon}) we obtain a Hamiltonian that is 
expressible in terms of the conserved charges, thus establishing integrability. We also undertake 
the algebraic Bethe ansatz to derive the associated exact solution. 

The integrable Hamiltonian that we will derive, and a particular case of the ones given in \cite{ddep,duk}, 
both belong to the following class of Hamiltonians:         
      
\bea
H=U N_b^2 +  \o N_b +  \sum_{j=1}^{M} \ve_j n_j+ 
 \sum_{j=1}^{M} g_j ( b^\dag {\cal S}_j^- + b {\cal S} _j^+ )
\label{ham}
\eea
where the $\ve_j$ denote $M$ two-fold degenerate energy levels for a system of fermions, $\o$ is the single energy level 
for condensed molecular bosons, $U$ is the interaction energy associated with S-wave scattering 
of the molecular bosons, and $g_j$ are the interaction strengths for scattering between Cooper pairs and molecular bosons.
In this article, the spin operators represent  Cooper pairs through a realisation in terms of canonical Fermi operators
$ {\cal S} _j^-= c_{j-}c_{j+},~ {\cal S} _j^+= c^\dag_{j+} c^\dag_{j-}, {\cal S}_j^z=(n_{j-}+n_{j+}  -1)/2  $
where $n_{j+}=c^\dag_{j+}c_{j+},\, 
n_{j-}=c^\dag_{j-} c_{j-}$ such that the following $su(2)$ relations are satisfied: 
$$ [{\cal S}^-_j, {\cal S} _j^+] = 2{\cal S} _j^z,~~~    [{\cal S}^z_j, {\cal S} _j^\pm] = \pm{\cal S} _j^\pm .  $$
Further, $b,b^\dag$ are canonical Bose operators obeying $[b,b^\dag]=1$.  We adopt  
the notation $N_b=b^\dag b$, $n_j = n_{j-}+n_{j+} $, $N_j = n_{j +} n_{j-}$. It can 
be verified that the total ``bosonic'' particle number (i.e. the number of Cooper pairs plus the number of molecular bosons)  
$N=N_b + \displaystyle\sum^{M}_{j=1}  N_j$
commutes with the Hamiltonian and is thus conserved.  
We note that for any $j$ the change of variable
\begin{equation}
g_j\rightarrow -g_j \label{unitary}
\end{equation} 
is a unitary transformation which is induced by the automorphism 
$$
{\cal S}_j \rightarrow -{\cal S}_j,~~ {\cal S}_j^+ \rightarrow -{\cal S}_j^+,~~ {\cal S}_j^z \rightarrow {\cal S}_j^z.
$$ 
Similar to BCS models \cite{dps,dl,jon} the Hamiltonian (\ref{ham}) displays Pauli blocking. 
For any unpaired fermion at energy level $\ve_j$ the action of the interactions with coupling $g_j$ is zero. 
This means that the Hilbert space can be decoupled into a tensor product of paired and 
unpaired fermion states for which the action of the Hamiltonian on the subsystem of 
unpaired fermions is automatically diagonal in the natural Fock basis. 
Below we let ${\cal I}$ denote the index set for the unblocked levels and let 
${\cal L}\leq M$ denote the cardinality of 
${\cal I}$. 

It was shown in \cite{ddep,duk} that 
\begin{equation}
U=0, ~~~~ g_j=g\,~~\forall\,j
\label{manifold1}
\end{equation}
 is an integrable manifold for the Hamiltonian (\ref{ham}). For this case  the energy eigenvalues are given by 
\begin{eqnarray*}
E&=&g\sum_{j=1}^N x_j   +\sum_{k\notin {\cal I}} \ve_k 
\end{eqnarray*}
where the parameters $x_j$ are solutions of the Bethe ansatz equations
\begin{eqnarray}
\frac{\omega}{2g}- \frac{1}{2}x_j -\frac{1}{4}\sum_{k\in {\cal I}}\frac{2}{2\varepsilon_k/g-x_j}
=-\sum^N_{k\neq j,\, k=1}\frac{1}{x_k-x_j}. 
\label{ddepbae}
\end{eqnarray} 
In the first part of the paper we will show that there is a second manifold of integrability which holds when 
\bea
\ve_j =\frac{ g_j^2}{2U}, ~~~~\frac{\o}{U} =( {\cal L} - 2N)
\label{manifold2}
\eea
and we will derive the corresponding Bethe ansatz solution. 
We remark at this point that there are two types of elementary excitations for the model, 
those for which Cooper pairs are broken and those for which they are not. 
Clearly for a non-pair breaking excitation ${\cal L}-2N$ remains constant. For a 
pair breaking excitation $N$ decreases by one and ${\cal L}$ decreases by two, 
such that ${\cal L}-2N$ remains constant. Hence the coupling constraints defined by 
(\ref{manifold2}) are independent of the type of excitation.  
 
In the second part of the paper we will show that for particular
submanifolds of  
both (\ref{manifold1}) and (\ref{manifold2}) there exists in each case
a mapping of the energy  
spectrum into part of the spectrum of a quasi-exactly solvable (QES)
Schr\"odinger equation. Quasi-exactly solvable problems are
characterised by having a 
finite number of wave functions and eigenvalues that can be found 
algebraically \cite{tur,ush}. The submanifolds of 
(\ref{manifold1}) and (\ref{manifold2}) we consider map onto the QES
sectors of  two Schr\"odinger equations.  The case of (\ref{manifold1}) is particularly
interesting because it naturally maps onto a 
$\PT$-symmetric \cite{BB,BBN} Schr\"odinger problem.


\section{Boundary QISM and construction of the Hamiltonian}


We will construct a family of commuting transfer matrices $t(u)$ via the boundary QISM \cite{skyl} from which the Hamiltonian will be obtained.
For the operator $t(u)$ acting on a Hilbert space of physical states, we require
$$[t(u),t(v)]=0, ~~\forall u,v\in {\mathbb{C}}. $$
A consequence of constructing the transfer matrix in this way is that by taking the series expansion 
$$ t(u) = \sum _{k=-\infty}^\infty t_k u^k$$
we have
$$[t_k, t_j]=0, ~~ \forall \,k,j,$$
which represent constants of the motion.
Below we demonstrate how the Hamiltonian is expressible as a function 
of these constants of motion and as a consequence deduce that the resulting
model is integrable. 

We begin with the Yang-Baxter equation
$$R_{12}(u-v) R_{13}(u-w) R_{23}(v-w) =R_{23}(v-w) R_{13}(u-w) R_{12}(u-v) $$
and use the $su(2)$ invariant $R$-matrix solution which has the form
\bea
R(u)= \pmatrix    {1& 0 & 0& 0  \cr
                        0& b(u) & c(u)& 0  \cr
                        0& c(u) & b(u)& 0  \cr
                 0& 0 & 0& 1  \cr
} \,,\nn
\eea
with $b(u)=u/{u+\eta},~~~c(u)=\eta/{u+\eta}$ where $\eta$ is an arbitrary complex parameter.
We require an $L$-operator, a realisation of the Yang--Baxter algebra acting on local physical spaces,
which here is given by a spin 1/2 realisation of the $su(2)$ Lie algebra  
\bea
  L_j(u)&=& I + \frac { \eta}{u } S_j \label{l}
\eea
for 
\bea
S_j= \pmatrix    {N_j-\frac 12 & {\cal S}_j^- \cr
                         {\cal S}_j^+ & \frac 12-N_j\cr } \,.\nn
\eea
We note that for 
$$
  \hat{L}_j (u)= I - \frac { \eta}{u-\eta } S_j  ,  \nn
$$
we have 
\begin{equation} 
 L_j(u) \hat{L}_j (u)=\lt(1- \frac {3 \eta^2}{4u(u-\eta) }\rt) I .
\label{linverse}
\end{equation}
The $L$-operator so defined satisfies 
\bea
R_{12}(u-v) L_{1}(u) L_{2}(v) =L_{2}(v) L_{1}(u) R_{12}(u-v). 
\label{ybel}
\eea
We also take the following realisation of the Yang--Baxter algebra in terms of the molecular boson operators $b,\,b^\dag$ 
\bea 
J(u)&=& \pmatrix{1+ \eta u+\eta^2 (N_b+1) &  \eta b  \cr 
 \eta b^\dag & 1   \cr}  \,,\nn\\
\hat{J}(u)&=& \pmatrix{1& - \eta b \cr
                                   - \eta b^\dag & 1+ \eta u+\eta^2 N_b  \cr} , \nn 
\eea
where we also have 
\begin{equation} 
J(u) \hat{J} (u)=(1+ \eta u) I. 
\label{jinverse}
\end{equation} 
The operator $J(u)$ obeys the following relation
\bea
R_{12}(u-v) J_{1}(u) J_{2}(v) =J_{2}(v) J_{1}(u) R_{12}(u-v). 
\label{ybej}
\eea
For further details on these realisations of the Yang--Baxter algebra, the reader is referred to \cite{jon}.
It will be useful to define
\beq
X=\pmatrix{ \eta (N_b+1)& - b(1+\eta^2 N_b ) \cr
                                    b^\dag & -\eta N_b \cr}  .\nn
\eeq

We construct the doubled monodromy matrix \cite{skyl} in terms of the local operators for Cooper pairs and 
molecular bosons. For a given index set ${\cal I}$ for the unblocked levels with cardinality ${\cal L}$, 
we first perform a relabelling such that 
${\cal I}\rightarrow {\cal I}'=\{1,2,\dots,{\cal L}-1,{\cal L}\}$. The doubled monodromy matrix is given by 
\bea 
T(u) &=& L_1(u-\e_1) \dots L_{\cal{L}}(u-\e_{\cal L}) J(u) K \hat{J}(-u)  \hat{L}_{\cal{L}} (-\e_{\cal L}-u) \dots \hat{L}_1 (-\e_1-u)  
\label{T}
\eea
where 
\bea
K =
\pmatrix    {1& 0 \cr
                        0& -1\cr } \,\nn
\eea
is the boundary $K$-matrix.  Note that here we will choose the
boundary $K$-matrices, usually denoted $K_+,~ K_-$, to be equal. 
These are not the most general $K$-matrices that can be chosen, but this choice is sufficient for 
the considerations of the present work. The $K$-matrix satisfies the equation 
\begin{equation}
R_{12}(u-v)(K\otimes I)R_{12}(u+v)(I\otimes K)
=(K\otimes I)R_{12}(u+v)(K\otimes I) R_{12}(u-v).  
\label{reflection}
\end{equation}  
The contribution $J(u) K \hat{J}(-u)$ to the doubled monodromy matrix is a dressing of the boundary 
$K$-matrix, which itself can be considered as an operator valued boundary $K$-matrix. We remark that operator valued boundary $K$-matrices have been used previously for the purpose of embedding Kondo impurities in integrable 
one-dimensional $t-J$ models \cite{zglg,frahm}. 

{}From (\ref{linverse},\ref{ybel},\ref{jinverse},\ref{ybej},\ref{reflection}) it follows that the doubled monodromy matrix satisfies
\begin{equation}
R_{12}(u-v)(T(u)\otimes I)R_{12}(u+v)(I\otimes T(v))=(I\otimes T(v))R_{12}(u+v)(T(u)\otimes I) R_{12}(u-v).  
\label{oybe}
\end{equation} 
Taking the expansion of the doubled monodromy matrix in inverse powers of $u$, we can express it as
\bea
T(u) 
&\approx&  u T_0 + u^0 T_1 + u^{-1}T_2 + u^{-2}T_3\nn
\eea
where
\bea
T_0&=& \eta I  \nn\\
T_1 &=& K + 2 \eta X + 2\eta^2 \sum_j^{\cal L}  S_j\nn\\
T_2 &=& 
\eta \sum_i^{\cal L}  (S_i K + K S_i )+ 2 \eta^2  \sum_i^{\cal L}  (S_i X + XS_i)-\eta^3 \sum_i^{\cal L}  S_i
+ \eta^3\left( \sum_{j<i}^{\cal L}  S_j S_i +\sum_{j>i}^{\cal L}  S_j S_i   +\sum_{i,j}^{\cal L}  S_j S_i  \right)\nn\\
T_3 &=&\eta\sum_i^{\cal L}  \e_i ( S_i K- K S_i) +2 \eta^2\sum_i^{\cal L}  \e_i (S_i X  -X S_i)
+ \eta^2 \left( \sum_{j<i}^{\cal L}  S_j S_i K + \sum_{j>i}^{\cal L}  K S_j S_i   + \sum_{i,j}^{\cal L}  S_j K S_i \right)
\nn\\
&&+ 2\eta^2 \sum_i^{\cal L}  \e_i^2 S_i -\eta^2 \sum_i^{\cal L}  K S_i+ O(\eta^3).
\nn
\eea
The transfer matrix is given by
\bea
t(u)&=& {\rm Tr} [K T(u)]\label{t}
\eea
which provides a family of commuting matrices:
$$[t(u),\,t(v)]=0~~~~\forall\,u,v. $$
A series expansion of the transfer matrix provides a set of mutually commuting operators,
\bea
t(u) &\approx & u t_0 + u^0 t_1 + u^{-1}t_2 + u^{-2}t_3, \nn
\eea
$$
[t_k , t_j ] = 0. 
$$
We find
\bea
t_0&=& 0 \nn\\
t_1 &=& 2+ 2 \eta^2(2N_b+1) + 2\eta^2 \sum_j^{\cal L}  (2N_j-1)\nn\\
t_2& =& 0 
\nn\\
t_3 &= & 4 \eta^2\sum_j^{\cal L}  \e_j ({\cal S}^-_j b^\dag + {\cal S}_j^+ b )+ 2\eta^2 \sum_i^{\cal L}  \e_i^2   (2N_i-1)  
+ \eta^2 \sum_{j<i}^{\cal L} \left[ 2(N_j - \frac 12 ) (N_i - \frac 12) +{\cal S}^-_j{\cal S} _i^+ + {\cal S}^+_j {\cal S}_i^-\right] \nn\\
&&+ \eta^2 \sum_{j>i}^{\cal L} \left[ 2(N_j - \frac 12 ) (N_i - \frac 12) +{\cal S}^-_j{\cal S} _i^+ + {\cal S} ^+_j {\cal S}^-_i\right] 
+ \eta^2 \sum_{j,i}^{\cal L} \left[ 2(N_j - \frac 12 ) (N_i - \frac 12) -{\cal S}^-_j{\cal S}_i^+ - {\cal S} ^+_j {\cal S} ^-_i\right] 
\nn\\
&&+O(\eta^3).
\nn
\eea
The Hamiltonian (\ref{ham}) is related to $t_3$ by
\begin{equation}
H=\lim_{\eta\rightarrow 0} \frac U{4\eta^2}\left( t_3  + 2 \eta^2\sum_{i=1}^{\cal L}  \e_i^2 
-4 \eta^2 N(N-{\cal{L}})- \frac{\eta^2 {\cal{L}}(2 {\cal{L}} -3)}2 \right) 
\label{lim} 
\end{equation}
after making the substitutions 
\bea
\ve_i = \frac{ U\e_i^2}2, ~~~ g_i = U{\e_i},~~~\o =U( {\cal L} - 2N)
\label{constraint}
\eea
where $i\in{\cal I}'=\{1,\dots {\cal L}\}$ label the unblocked levels. The above conditions may be easily relabelled in terms of the original index set ${\cal I}$.

 \section{ Algebraic Bethe ansatz solution \label{aba}}

Sklyanin's boundary QISM for  the doubled monodromy matrix \cite{skyl} 
was adapted for the Gaudin magnet in \cite{hik,hik1} to obtain a solution using the algebraic Bethe ansatz (ABA) method.  
Using this general scheme, we outline the boundary ABA to develop a solution for the Hamiltonian (\ref{ham})
subject to the constraints (\ref{manifold2}). 

With the above definition for the transfer matrix  (\ref{t}),  we solve the eigenvalue problem
$$t(u)  \Psi=\L(u) \Psi$$
by the boundary ABA method.
Representing the doubled monodromy matrix as
\beq
T (u)= \pmatrix{   A(u) &  B(u) \cr
                                    C(u) & D(u) \cr}  \,,
\eeq
the transfer matrix is given by
$$t(u) \Psi= (A(u) - D(u) ) \Psi.$$
The relations resulting from (\ref{oybe}) are quite complicated in terms of which contributions are relevant to the diagonalisation 
of the transfer matrix.  For this reason, we define a new $\hat A(u)$ through 
$$\hat{A}(u)=(2u+\eta) A(u) - \eta D(u) .$$
We can now rewrite the relevant relations arising from (\ref{oybe}) in a more convenient form:
\bea
\hat{A} (u) C(v) &=&  \frac {(u-v+\eta)(u+v+2\eta)}{(u+v+\eta)(u-v)} C(v) \hat{A}(u) - 
\frac  {2\eta(u+\eta)}{ (2v+\eta) (u-v)}C(u) \hat{A}(v)\nn\\
&+& \frac  {4v\eta(u+\eta)}{(2v+\eta)(u+v+\eta)}C(u) D(v)\\ 
D(u) C(v) &=&  \frac  {(u+v)(u-v-\eta)}{(u+v+\eta)(u-v)} C(v)D (u)
+ \frac  {2v\eta}{(2v+\eta)(u-v)}C(u) D (v) \nn\\
&-&
\frac  {\eta }{(u+v+\eta)(2v+\eta)} C(u) \hat{A}(v).
\eea
Correspondingly, the transfer matrix may now be expressed as
$$t(u)= \frac 1{2u+\eta}\hat{A}(u) - \frac {2u} {2u+\eta}{D}(u).$$

For the construction of the eigenstates, the reference state is chosen to be 
\bea
|\Omega\re = |0\re \bigotimes_{i=1}^{\cal L}  
\pmatrix{   0\cr
     1\cr}_i 
\eea
such that $B(u)|\Omega\re=0$. 
We find that the action of $\hat {A}(u),~ D(u)$ on this reference state is
\bea
\hat{A}(u) |\Omega\re &=& 2(u+\eta) \a(u) \hat{\a} (-u) |\Omega\re \\
{D}(u) |\Omega\re &= & -\d(u) \hat{\d}  (-u)  |\Omega\re 
\eea
where
\bea
\a(u) |\Omega\re &= & (1+\eta u+ \eta^2 )\prod _{i=1} ^{\cal{L}}\left(1 - \frac {\eta}{2(u-\e_i)}\right)   |\Omega\re  \nn\\
\hat{\a} (-u)|\Omega\re &= &  \prod_{i=1} ^{\cal{L}}\left(1 - \frac {\eta}{2(u+\e_i)}\right)    |\Omega\re \nn\\
\d(u) |\Omega\re &= & \prod_{i=1} ^{\cal{L}} \left(1 + \frac {\eta}{2(u-\e_i)}\right)     |\Omega\re \nn\\
\hat{\d}  (-u)  |\Omega\re &= &  (1-\eta u) \prod_{i=1} ^{\cal{L}} \left(1 + \frac {\eta}{2(u+\e_i)}\right)     |\Omega\re .\nn
\eea
Taking the Bethe ansatz states to be a product of the creation operators on the reference state,
$$  \Psi = \prod_{\a=1}^{N} C(v_\a) |\Omega\re,  $$
we find that for 
$$ t(u)  \Psi= \L(u)  \Psi+ \mbox{``unwanted terms"}$$
the co-efficient $\L(u)$ is given by 
\bea
\L(u)& =& \frac {2(u+\eta)(1+\eta u +\eta^2)}{2u+\eta} \prod^N_{\a=1}    \frac {(u-v_\a+\eta)(u+v_\a+2\eta)}{(u+v_\a+\eta)(u-v_\a)} 
\prod^{\cal L}_{i=1}  \frac {  (u-\e_i   - \eta/2 ) (u+\e_i   +\eta/2 ) }{ (u-\e_i) (u+\e_i+\eta) }  \nn\\
&+& \frac {2u(1-\eta u )}{2u+\eta} \prod^N_{\a=1}    \frac {(u+v_\a)(u-v_\a-\eta)}{(u+v_\a+\eta)(u-v_\a)} 
\prod^{\cal L}_{i=1}   \frac{ ( u-\e_i  +\eta/2) ( u+\e_i  +3\eta/2)    }{  (u-\e_i) (u+\e_i+\eta)  } .  \label{lam}  
\eea
Cancellation of the unwanted terms above, which is equivalent to imposing that $\L(u)$
has no poles,  leads to the Bethe ansatz equations
\bea
\frac {v_\b (1-\eta v_\b)}{(v_\b+\eta)(1+\eta v_\b +\eta^2)}   
\prod^{\cal L}_{i=1}   \frac { (v_\b+\e_i + 3\eta/2) ( v_\b-\e_i  +\eta/2)      }{(v_\b+\e_i   + \eta /2) (v_\b-\e_i  - \eta/2 )  }   
&=&
\prod^N_{\a\neq \b, \a=1}  \frac  {(v_\b-v_\a+\eta)(v_\b+v_\a+2\eta)} 
{(v_\b+v_\a )(v_\b-v_\a-\eta)},  \nn\\
&&~~~~~~~~~~~~~~ ~~~\b=1\dots N.    \label{bae1}
\eea

 \subsection{ Results in the quasi-classical limit \label{qcl}}

Expanding the eigenvalue of the transfer matrix (\ref{lam}) in inverse powers of $u$ we obtain
\bea
\L(u) &\approx & 
-2\eta^2 {\cal{L}}+4\eta^2N+2+2\eta^2 +\frac {\eta^2}{u^2}
\left[{\cal{L}}^2-4{\cal{L}}N-2\sum_{i=1}^{\cal L} \e_i^2  + 4N^2-\frac{3\cal{L}}2 +4 \sum_{\a=1}^N  v_\a^2 \right]
\nn\\
&&= \l_0 u + \l_1 u^0 + \l_2 u^{-1}  + \l_3 u^{-2}  \nn  
\eea
and in particular
\bea
\l_3 &=& \eta^2 \left[4 \sum_{\a=1}^N   v_\a^2-2\sum_{i=1}^{\cal L}   \e_i^2  + 4N^2 
-4{\cal{L}}N + {\cal{L}}^2-\frac {3{\cal{L}}}2 \right]. \nn
\eea
In view of (\ref{lim}) this leads us to the energy of the Hamiltonian being  
$$E=  U\sum_{\a=1}^N   v_\a^2 +\sum_{k\notin {\cal I}} \ve_k  $$
where ${\cal I}$ again denotes the index set for the unblocked levels.  
The corresponding Bethe ansatz equations arising from (\ref{bae1}) in the quasi-classical limit are 
\bea
  - 1  - \frac 1 {2v_\b^2} + 
   \sum_{k\in {\cal I} } \frac {1}{ (v_\b^2-2\ve_k/U)} 
&=&
   \sum_{\a\neq \b, \a=1}^N \frac  { 2}{(v_\b^2-v_\a^2 ) }. \label{qcbae}
\eea

\section{Mapping to a Schr\"odinger equation}


For the remainder of the paper we will restrict to a degenerate case for 
which $\ve_j$ is independent of $j$ and consider
only non-pair breaking excitations such that ${\cal L}=M$. For both exact solutions given by the Bethe ansatz equations (BAE) (\ref{ddepbae},\ref{qcbae}), 
we will demonstrate that the spectrum of the model, with a suitable scaling, can be mapped into that of a one-dimensional Schr\"odinger equation.

\subsection{A first mapping}
We begin with the
 BAE (\ref{qcbae}) derived in the previous section. First observe through (\ref{constraint}) that the variable $U$ determines
the overall scaling of the model.   
We set $U=-1$ (so the S-wave scattering is attractive) and also 
$$~-2\ve_j=\g, \forall j,  ~~~~\quad v_\b^2= x_\b.$$  
The BAE are now
\bea 
-1 -\frac 1 {2 x_\b}+\frac{{ M }}{x_\b-\g}=\sum_{\a\neq \b,\a=1}^N\frac{2}{x_\b-x_\a}. \nn
\eea 
It is convenient to shift the roots, $x_\b \rightarrow x_\b + \g/2$.  Then the BAE become
\beq 
-1 - \frac {1} {2({x_\b + \g/2})}+\frac{ M }{x_\b-\g/2}=\sum_{\a\neq \b,\a=1}^N\frac{2}{x_\b-x_\a}. \label {bae}
\eeq 
Following the procedures taken in \cite{rktb,zlm}, we choose the following ansatz for an ordinary differential 
equation (ODE):
\bea
F(y) = [y^2 - \lt( \g/2\rt)^2] Q''(y) - \lt[ -\frac 12 (y- \g/2) + { M }(y+ \g/2) - y^2 + \lt(\g/2\rt)^2 \rt] Q'(y).
\label{ode1}
\eea
For $\displaystyle Q(y) = \prod^N_{\b=1} (y - x_\b)$, we can check
that $F(x_\b)=0$ using the BAE (\ref{bae}) and noting  
\bea 
Q'(y)&=&Q(y) \sum_{\a=1}^N\frac{1}{y-x_\a} \nn \\
Q''(y)&=&Q(y)\sum_{\b=1}^N\sum_{\a\neq \b,\a=1}^N\frac{1}{(y-x_\b)(y-x_\a)}. 
\nn 
\eea 
The degree of the polynomial $Q(y)$ is $N$, so in the RHS of (\ref{ode1}) the highest order term is
$N+1$. Thus we can fix $F(y)$ to be
\bea
F(y) = ( A y + B) Q(y).\nn
\eea
It can be shown that the leading order term gives 
$$A = N.$$

Next consider
\bea 
\frac {Q'(\g/2)}{Q(\g/2) }  &=& - \sum_{\a=1}^N\frac{1}{x_\a-\g/2} \nn \\
\frac {Q'(-\g/2)}{Q(-\g/2) }  &=& - \sum_{\a=1}^N\frac{1}{x_\a+\g/2}. \nn 
\eea 
{}From the ODE (\ref{ode1}), we also see that
\bea 
\frac {Q'(\g/2)}{Q(\g/2) }  &=& -\frac N{2{ M }} - \frac B{{ M }\g}\nn \\
\frac {Q'(-\g/2)}{Q(-\g/2) }  &=& N - \frac {2B}\g .\nn 
\eea 
Rearranging for $B$ we find
\bea
B &=& - \frac {{ M }\g}2 \frac {Q'(\g/2)}{Q(\g/2) }  - \frac \g 4 \frac {Q'(-\g/2)}{Q(-\g/2) }  \nn\\
&=&  \frac {{ M }\g}2 \sum_{\b=1}^N\frac{1}{x_\b-\g/2}  + \frac \g 4  \sum_{\b=1}^N\frac{1}{x_\b+\g/2}. \nn
\eea
We can simplify $B$ using an identity from the BAE.  Taking the BAE (\ref{bae}), multiplying by $x_\b$ and taking the sum over $\b$  
gives us
\bea
-\sum_{\b=1}^N x_\b + (-1/2 + { M })N + \frac \g 4 \sum_{\b=1}^N\frac{1}{x_\b+\g/2} 
+ \frac {{ M }\g}2 \sum_{\b=1}^N\frac{1}{x_\b-\g/2}  =  N(N-1).\label{id4} 
\eea
Using this identity, we obtain $B$ as follows
$$B = -E - N\lt({ M }-N+ \frac 12\rt) - \frac {N\g}2 , $$
where for the energy we have
$$ E = -\sum_{\b=1}^N  v_\b^2 =- \frac {N\g}2 - \sum_{\b=1}^N x_\b.$$
Hence $Q(y)$ satisfies the differential equation
\bea
\lt[Ny - E \rt. &-& \lt. N\lt( { M }-N+\frac 12+\frac \g 2\rt)\rt]Q(y) = (y^2 - \lt( \g/2\rt)^2) Q''(y) \nn\\
&-& \lt[ -\frac 12 (y- \g/2) + { M }(y+ \g/2) - y^2 + \lt(\g/2\rt)^2 \rt] Q'(y).
\label{odeme}
\eea

Now we look to put the ODE (\ref{odeme}) into the form of a
Schr\"odinger equation. 
We start with the substitution $y= -  (\g  \cosh x)/2 $, so that
\bea 
\frac{dQ}{dx} &=& -\frac \g 2 \sinh (x) \frac{dQ}{dy}, \nn \\
\frac{d^2Q}{dy^2} &=& \frac {4}{{\g^2}\sinh^2(x)} \lt(\frac{d^2 Q}{dx^2} - \coth(x) \frac{dQ}{dx}\rt). \nn  
\eea 
Now (\ref{odeme}) becomes   
\bea
\frac{d^2 Q}{dx^2} -\frac { \sinh (x)}{2(\cosh(x)+1) }  \lt[ 1 +2{ M } 
 + {\g} (\cosh ( x)+1) \rt] \frac{dQ}{dx} \nn\\
=\lt[- \frac {N\g} 2 (\cosh (x)+1)  -  E - N\lt( { M }-N+\frac 12\rt)\rt]Q(x)   
.\nn
\label{ode2}
\eea
Next we put 
$$\Psi(x)= \exp (-g(x)) Q(x)  $$ 
and we need to find $g(x)$ such that upon substituting $\Psi(x)$ into
the ODE (\ref{ode1}), the $\Psi '(x)$ term cancels. (Note that 
the $\Psi(x)$ here is not related to the state vector $\Psi$
introduced  earlier in the ABA section.) 
For
\bea
Q'(x) &=& \exp(g(x)) [g'(x) \Psi (x)+  \Psi'(x)] \nn\\
Q''(x) &=& \exp(g(x)) [ ( g''(x) + g'(x)^2 ) \Psi (x) + 2 g'(x) \Psi'(x)+ \Psi''(x) ],\nn
\eea
we find that the contribution to $\Psi'(x)$ (which we need to eliminate) is as follows:
$$ g'(x) =\frac { \sinh (x)}{4(\cosh(x)+1) }  \lt[ 1 +2{ M } 
 + {\g} (\cosh ( x)+1) \rt]. $$
This is easily solved for $g(x)$:
$$ g(x) = \frac 14[ \g \cosh (x) + (2{ M } +1) \ln ( \cosh (x) + 1) ].$$
We now obtain
$$\Psi(x)= (\cosh \lt(x\rt)+1)^{-({ M }/2 +1/4) } \exp \lt( -\frac \g4 \cosh (x) \rt)     Q(y)  $$  
and also note $$ g''(x) = \frac {\g \cosh (x) ( 1 + \cosh (x) ) + 2{ M } + 1} { 4 ( \cosh (x) + 1)} .$$

Next we can rewrite the ODE (\ref{ode1}) as a Schr\"odinger equation
\bea
- \Psi''(x) + V(x) \Psi(x) = E \Psi(x)\nn
\eea
where 
\bea
V(x)
&=&-N\left(\frac{\g}{2}(\cosh (x)+1)+{ M }-N+\frac{1}{2}\right) 
+\frac{(2{ M }+1)^2}{16}-\frac{(2{ M }+1)\g}{8} \\ 
&&~~~~ -\frac{(2{ M }+1)(2{ M }+3)}{8(\cosh (x)+1)} 
 +\frac{(2{ M }-1)\g\cosh (x)}{8}+\frac{\g^2\sinh^2(x)}{16},\nn
\eea
and
\bea
\Psi (x) &=& ( \cosh \lt(x\rt)+1)^{-({ M }/2 +1/4) } \exp \lt( -\frac \g4 \cosh (x) \rt) 
\prod_{\a=1}^N \lt( \frac \g 2 \cosh x + x_\a\rt).
\eea 
This potential belongs to a class of known quasi-exactly solvable
potentials \cite{tur,qes2}.

\subsection{A second mapping}


We can perform a similar mapping for the case of the solution (\ref{ddepbae}). 
Setting $\ve_j = \e, ~ g=-1$ and performing a shift in the roots $x_j\rightarrow x_j +\e$, the BAE are 
\bea 
 2\e-\o - x_j +
  \frac { M }{ x_j} 
=-
   \sum_{k\neq j, k=1}^N \frac  { 2}{(x_k-x_j ) } \nn
\label{bae2}
\eea
where the energy is given by 
\bea
 E = - \sum_{j=1}^N x_j - 2 \e N.
\eea 
We choose the following ansatz for the ODE:
\bea
F(y) = y Q''(y) +  \lt[ (\o-2\e ) y + y^2 -  { M } \rt] Q'(y).
\label{oded1}
\eea
For $Q(y) = \displaystyle\prod^{N}_{j=1}  (y - x_j)$, we can easily
check that $F(x_j)=0$ using the BAE.  Now set 
\bea
F(y) = ( A y + B) Q(y).
\label{f}
\eea
It can be shown that the leading order term gives 
$$A = {N} .$$
Next consider
\bea 
\frac {Q'( 0)}{Q( 0) }  &=& - \sum_{j=1}^N\frac{1}{x_j} \nn \\
\frac {Q''( 0)}{Q( 0) }  &=&  \sum_{j=1}^N \sum _{k \leq j,k=1}^N \frac{1}{x_j x_k} .\nn 
\eea 
Also 
$$F(0) = B = -{ M } Q'(0) = { M } \sum_{j=1}^N \frac 1 {x_j}.$$
We simplify $B$ using an identity from the BAE.  Taking the summation over $j$ in the BAE, 
we obtain $B$ as follows
$$B = N \o-E.$$
Hence $Q(y)$ satisfies the differential equation
\bea
 y Q''(y)+ \lt[(\o -2\e )  y + y^2 -  {{ M }} \rt] Q'(y) + \lt[-{N} y -N \o+E\rt]Q(y)  =0.
\label{oded2}
\eea
We make the substitution $y=  {x^2}/4$ so that
\bea 
\frac{dQ}{dx} &=& \frac x2 \frac{dQ}{dy}, \nn \\
\frac{d^2 Q}{dy^2} 
&=& -\frac 4{x^3} \frac{d Q}{dx} +\frac 4 {x^2} \frac{d^2Q}{dx^2}. \nn 
\eea 
Now the ODE (\ref{oded2}) becomes   
\bea
&&\lt[ -  \frac {Nx^2}4 -N \o+E  \rt]Q(x) +  
 \frac{d^2 Q}{dx^2}  + \lt[ \frac{x^3}8  - \frac {2{ M } +1}x- \frac { (\o-2\e )x}{2} \rt] \frac{dQ}{dx}=0 .\nn
\eea

Next we put 
$$\Psi(x)= \exp (-g(x)) Q(x)  $$ 
and substituting $\Psi(x)$ into the ODE,
the contribution to $\Psi'(x)$ is given by 
$$ 2 g'(x) = \lt[ - \frac {x^3} 8 + \frac {(2{ M }+ 1) } x +\frac {(\o-2\e )x}{2} \rt],$$
which we solve for $g(x)$ to obtain
$$ g(x) =  \frac{ (2{ M } + 1) }2 \ln x - \frac {x^4} { 64}- \frac {x^2 (\o-2\e ) }{8}  .$$
Setting $z=x/2,\,\beta= \o-2\e $, we can now rewrite the ODE (\ref{oded2})  as a Schr\"odinger equation
\bea
- \Psi''(z) + V(z) \Psi(z) = \bar E \Psi(z),
\eea
where 

\bea 
V(z)&=&  z^6 +2\beta z^4 + \lt \{\beta^2 + 2(1+2N-M)  \rt\}z^2 
+ \frac{ (M+ 3/ 2 ) (M + 1 /2 )}{z^2} \label{sextic} \\
\bar E&=&-2\beta(2N-M)-4\sum_{j=1}^Nx_j , \nonumber \\  
\Psi (z) &=& z^{  {-(2{ M } +1)}/2 } \exp \lt( 
\frac {z^4}{4}+\frac {\beta z^2}{2}   \rt)     \prod_{\a=1}^{N} \lt ( {z^2}  - x_\a\rt)
\label{wf} 
\eea
and the parameters $x_j$ are roots of the BAE (\ref{bae2}). 

The quasi-exact solvability of the potential (\ref{sextic}) has previously been discussed in \cite{ush}  
(see Chapter 2.2). 
However some care is needed in order to properly embed the eigenfunctions
into a Hilbert space of states. Specifically, the 
difference 
between the standard Schr\"odinger problem and the one above 
lies in the boundary conditions imposed on the eigenfunctions
in order to obtain the Hilbert space. 
In the standard case the wave functions 
are required to  be square-integrable, and thus decay at $\pm \infty$ along the
real axis.  
The  wave function (\ref{wf}) 
is not square-integrable on the real axis,
since at large real $|z|$ it clearly blows up. However it is  
square-integrable if we instead define 
it on a contour 
which lies in wedges of the complex
plane of open angle $\pi/4$ centered about $\arg z = -\pi/4$ 
and $\arg z = -3\pi/4$, and 
distorted to pass below the origin since $(M{+}3/2 ) (M{+}1
/2 ) \neq 0$ \cite{BT}.  These are $\PT$-symmetric \cite{BB,BBN} boundary
conditions, and 
our QES potential (\ref{sextic}) 
leads to the angular-momentum generalisation  of the $\PT$-symmetric
Schr\"odinger equation studied in \cite{bm}.
It is interesting to note that 
 for (\ref{sextic}), the QES spectrum is necessarily real, since it has been derived from a mapping 
of a particular integrable manifold of the Hermitian Hamiltonian (\ref{ham}). By contrast, the potential studied in  
\cite{bm} admits a region of broken $\PT$-symmetry where the QES spectrum becomes complex.
 We always map to the region of unbroken $\PT$-symmetry because of the 
 constraint $M>0$.

Finally, we remark that 
the QES  potential (\ref{sextic}) has an interesting duality property
 under the transformation $\beta \to -\beta$, $z \to iz$. The quasi-exactly solvable sector of the
spectrum maps onto itself via $\bar E\to -\bar E$
\cite{shif,kuw} while the 
 non-QES sectors remain unrelated. In our picture this duality
 is a trivial consequence of the freedom we have to redefine the Bose
 operators $b \to -b
 $ and $b^{\dagger} \to -b^{\dagger}$.   If we send $\omega\to -\o$
 and $\e \to \e$ 
the Hamiltonian
becomes $-H$ and the energy eigenvalues  change sign.  Hence the QES
energy levels become $- \bar E$.


\section{Conclusion} 


The general Hamiltonian (\ref{ham})
has two distinct integrable manifolds in parameter space which are given by 
(\ref{manifold1},\ref{manifold2}) and exact Bethe ansatz solutions given by (\ref{ddepbae},\ref{qcbae})
respectively. 
For a certain submanifold in each instance, 
we have shown that the 
eigenspectrum and eigenstates can be mapped to the algebraic sector of a
QES Schr\"odinger potential. It is surprising that for one solution the 
mapping is to $\PT$-symmetric eigenstates defined on a contour in the complex plane, while for the other case the eigenstates are defined on the real axis. The implications of this curious result in relation to the different physical properties of the Hamiltonians warrant further investigation.

\vspace{.5cm}
{\bf Acknowledgements:}  
We thank the Australian Research Council for financial support.

\end{document}